# From Stability to Change: The Potential Application of Bifurcation Theory to Opinion Dynamics Considerations

Yasuko Kawahata †

Faculty of Sociology, Department of Media Sociology, Rikkyo University, 3-34-1 Nishi-Ikebukuro,Toshima-ku, Tokyo, 171-8501, JAPAN.
ykawahata@rikkyo.ac.jp,kawahata.lab3@damp.tottori-u.ac.jp

**Abstract:** This study proposes a nonlinear dynamics framework for understanding social interactions and behavioral patterns of online communities. Based on the work of Steven Strogatz, we apply the bifurcation phenomena, in particular pitchfork bifurcation, saddle node bifurcation, and transcritical bifurcation, to online access numbers and social media information diffusion patterns, and analyze how these bifurcations are associated with collective changes in opinion and the sudden We explore the possibility of analyzing how these bifurcations are related to collective changes in opinion and the sudden diffusion of trends.Starting from a basic text on nonlinear dynamics by Strogatz (1994), we integrate important contributions in social network theory through a review article by Strogatz (2001) on the structure of complex networks. Watts and Strogatz's (1998) work on small-world networks shows the relationship between local effects of collective action and overall dynamics, while Strogatz (2003) extensively explored synchronization phenomena in nature and human society.The paper draws on a detailed understanding of bifurcation theory by Gückenheimer and Holmes (1983), an exhaustive bibliography of applied bifurcation theory by Kuznetsov (2004), and an introduction to bifurcation theory and chaos theory by Wiggins (2003), exploring the potential applications each suggests for social phenomena. In addition, the practical approach to bifurcation theory offered by Seidel (2010) provides a better understanding of different types of bifurcations, including transcritical bifurcations in particular, and provides a foundation for using these concepts as a research method in the social sciences.This approach opens new avenues for studying the evolution of collective opinion online, the origin and diffusion of fads, and the transformation of behavior within digital ecosystems. Our model reflects the reality of contemporary digital communication, where the speed of information exchange is rapid and individual cognitive processes are complex, and aims to understand social dynamics. In recent years, nonlinear dynamical systems theory has become essential for analyzing complex systems that exhibit unpredictable behavior in the natural and social sciences. In particular, Melnikov's method has been established as a fundamental technique for evaluating the stability of Homoclinic and Heteroclinic trajectories in dynamical systems and for understanding the emergence of chaos. This has important implications for understanding how minute perturbations can cause a system to transition to chaotic motion. In this study, we use this mathematical framework to investigate specific applications in various fields.

## 1. Introduction

As the digital tapestry of global communication weaves increasingly complex patterns, the study of online behavior has evolved beyond mere social curiosity into an important scientific endeavor. The interplay of individual choices within the vast network of social media creates collective behaviors that reflect and shape our social norms. These phenomena resemble emergent patterns found in natural systems and are ripe for exploration through the mathematical elegance of branching theory. Bifurcation theory is a field heavily influenced by the insights of Steven Strogatz. We also propose to integrate theory and experience by leveraging insights from dynamic systems to analyze the rise and fall of digital interaction. Our work on bifurcation theory provides a solid foundation for this effort, with Guckenheimer and Holmes (1983) and a comprehensive look at applied bifurcation theory underlying important points of change.

### 1.1 Nonlinear Dynamics Theory

Beginning with his original 1994 work on nonlinear dynamics, Strogatz's pioneering work reached out to the complex ballet of systems that oscillate between order and chaos. His analysis of rake bifurcations provided a conceptual scaffold for understanding how the equilibrium of a system can diverge into multiple new pathways. In the realm of social media, this translates into viral moments where a single story splits into numerous competing storylines, each vying for dominance in the digital age spirit. 2001 review by Strogatz shed light on the subtle structures that underpin complex networks and provided a lens through which to view the interconnected world of online communities. Here, a phenomenon similar to the bifurcation of saddles and nodes was observed in the web of digital interaction, where the gradual accumulation of online discourse could suddenly change and develop into a cascading viral phenomenon or dissipate into the noise of the Internet.

## 1.2 Bifurcation Theory

This research aims to combine the robust framework of bifurcation theory with the hydrodynamics of online social trends. We draw on foundational theories developed by scholars such as Guckenheimer and Holmes (1983), whose work on dynamical systems provides in-depth knowledge for understanding change phenomena, and Kuznetsov (2004), whose work provides an in-depth exploration of bifurcation phenomena. In the applied setting. We stand at the intersection of order and chaos, and collective behavior on social media platforms oscillates between predictable patterns and abrupt changes.

### 1.2.1 Saddle-node Bifurcation

Define the saddle-node bifurcation function as:

$$f(r, x) = r + x^2,$$

where $r$ is the bifurcation parameter and $x$ is the state variable.

Create a grid for $r$ and $x$ using the meshgrid function:

$$R, X = \text{np.meshgrid}(r_{\text{values}}, x_{\text{values}}).$$

Values of the saddle-node bifurcation on the grid:

$$Z = f(R, X).$$

### 1.2.2 Pitchfork Bifurcation

Define the pitchfork bifurcation function $f : \mathbb{R} \times \mathbb{R} \to \mathbb{R}$ as:

$$f(r, x) = rx - x^3,$$

where $r$ is the bifurcation parameter and $x$ is the state variable.

Create a grid for $r$ and $x$ using the meshgrid function:

$$R, X = \text{np.meshgrid}(r_{\text{values}}, x_{\text{values}}).$$

Also values of the pitchfork bifurcation function over the grid:

$$Z = f(R, X).$$

### 1.2.3 Transcritical Bifurcation

Define the transcritical bifurcation function $f : \mathbb{R} \times \mathbb{R} \to \mathbb{R}$ as:

$$f(r, x) = rx - x^2,$$

where $r$ is the bifurcation parameter and $x$ is the state variable, too.

Create a grid for $r$ and $x$ using the meshgrid function:

$$R, X = \text{np.meshgrid}(r_{\text{values}}, x_{\text{values}}).$$

Also values of the transcritical bifurcation function over the grid:

$$Z = f(R, X).$$

### 1.2.4 Hopf Bifurcation

Set the parameters for the Hopf bifurcation, where $r > 0$ indicates that the fixed point is unstable and a limit cycle occurs:

$$r = 0.1, \quad b = 1.0$$

Define the Hopf bifurcation vector field function $f : \mathbb{R}^2 \to \mathbb{R}^2$ with parameters $r$ and $b$ as:

$$u(x, y) = rx - x^3 - y,$$
$$v(x, y) = x + by.$$

This system describes the evolution of the state variables $x$ and $y$.

Create a grid for the vector field using the meshgrid function:

$$X, Y = \text{np.meshgrid}(x_{\text{values}}, y_{\text{values}})$$

Compute the vector field $U, V$ on this grid:

$$U, V = \text{hopf\_bifurcation}(X, Y, r, b).$$

If $r > 0$, plot an approximation of the limit cycle:

$$\theta = \text{np.linspace}(0, 2\pi, 100), \quad \texttt{plt}(\sqrt{r}\cos(\theta), \sqrt{r}\sin(\theta),'r-')$$

### 1.2.5 Limit Cycle

Limit cycle is a situation in a dynamic system in which the behavior of the system settles into a periodic trajectory. In the context of consensus building, a phenomenon analogous to a limit cycle can be imagined as a situation in which opinions continue to fluctuate cyclically without complete convergence.

(1) **Political polarization**

Situation in which political opinion is divided into two extreme points, each supported for a period of time, and then swings to the opposite side. This cyclical fluctuation may indicate a state of affairs that becomes a permanent conflict and does not lead to a middle ground or consensus.

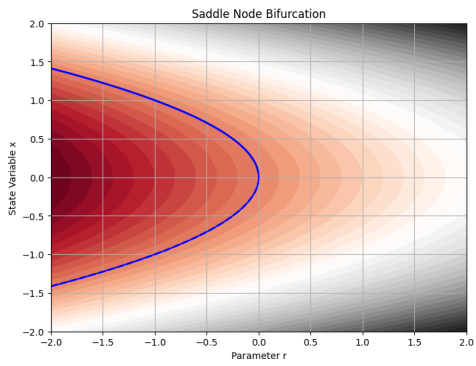

Fig. 1: Saddle Node Bifurcation

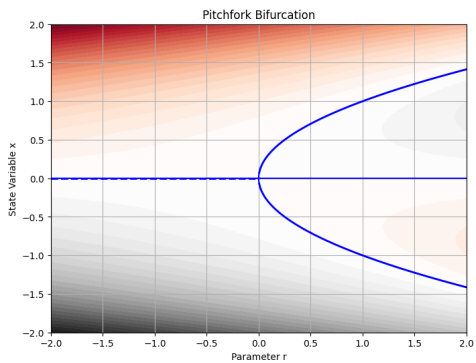

Fig. 2: Pitchfork Bifurcation

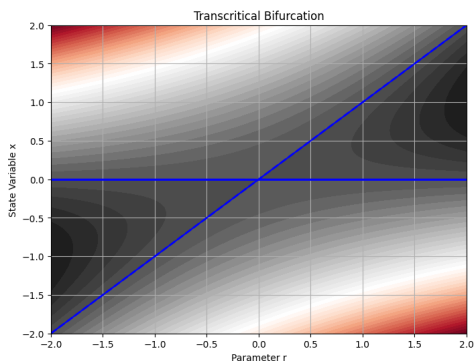

Fig. 3: Transcritical Bifurcation

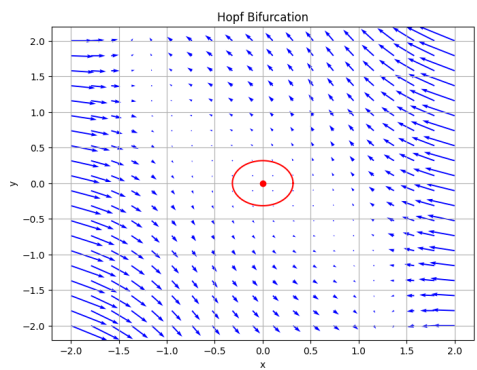

Fig. 4: Case Study:Hopf Bifurcation

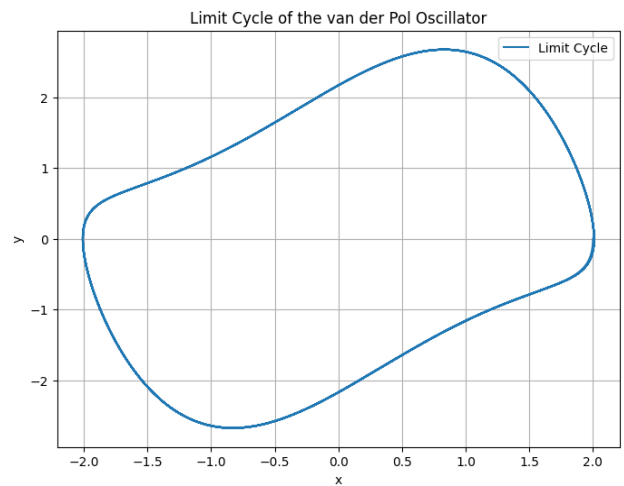

Fig. 5: Limit Cycle of the van der Pol Oscillator

(2) **Fashion and trend cycles.**
Phenomenon in which old styles become fashionable again in a certain pattern, and old and new trends switch periodically. This can be seen as a return to an old consensus (old trend) before a new consensus (new trend) is formed.

(3) **Economic business cycle**
Economy goes through cycles of expansion and contraction, where consensus among market participants (e.g., agreement on price levels) is temporarily achieved, only to fall back into disagreement. This disagreement appears periodically and may behave like a limit cycle.

(4) **Public Cycle of Opinions**
Situation in which public opinion on a particular topic or issue in the media or in public debate is such that one opinion becomes dominant while the opposite opinion gains support over time. This is sometimes referred to as a public cycle and illustrates how social consensus is not stable.

These examples suggest a situation in which consensus building has a certain pattern or cycle and does not lead to complete convergence. In actual social phenomena, these cyclical fluctuations are complicated by external factors and internal dynamics, so they do not coincide perfectly with mathematical limit cycles.

The realm of cyberspace is a living laboratory for observing the phenomena described by branching theory. Strogatz's foundational text, written in 1994, serves as a guidepost to guide us through the complexities of digital social structures through the explanatory power of rake bifurcations. These bifurcations, catalyzed by influential tweets and groundbreaking news articles, reverberate through the digital world as moments when the unified flow of online dialogue diverges into different ways of thinking.

The study also examines the heart of these fluctuations with the goal of uncovering the mechanisms that drive the formation, evolution, and in some cases dissolution of online consensus. Through a meticulous compilation of case studies and numerical simulations, we aim to create a narrative that not only elucidates the current state of digital communication, but also predicts its future trajectory. In a whirlwind of online interactions intersecting billions of digital footprints, the dynamics of social media engagement present a fascinating frontier for scientific research. The complex dance of likes, shares, and comments follows patterns that point to a deeper underlying structure. This structure will be governed by the same mathematical principles that Steven Strogatz revealed in his seminal contributions to the field of nonlinear dynamics.

Through this investigation, we seek not only to interpret the current state of online interaction, but also to predict the evolution of digital communities and prepare for the waves of change that may occur in the future of communication. This journey through the landscape of digital society, illuminated by the principles of bifurcation theory, aims to articulate the subtle but powerful forces shaping and driving change in the online world. In the ever-evolving realm of digital interaction, where humanity's transient thoughts and feelings are captured in a web of binary code, the study of online behavior transcends its early origins, psychology, sociology, mathematics and computer science. Once considered a mere reflection of the physical world, the digital realm is now recognized as a unique crucible of social phenomena governed by the principles of nonlinear dynamics and chaos theory, as detailed in Stephen Strogatz's book.

## 1.3 Mathematical Concepts Such as the Poincaré Cutting Plane

The application of mathematical concepts such as the Poincaré cutting plane of the Lorenz attractor to meta-social or discursive considerations may be devise as a metaphorical or philosophical approach. The Lorenz attractor is a classic example of chaos theory, representing a system that exhibits unpredictable yet deterministic behavior. Such properties can provide interesting parallels for considerations about the dynamics of society and discourse. When analyzing the Poincaré cutting plane of the Lorenz attractor in a sociological or discursive context, social phenomena and speech flows are often chaotic in nature. Small events can cause major social changes, while large events can have less unexpected effects. This "butterfly effect" suggests the difficulty of decision-making and forecasting in social contexts. The Lorenz attractor is a concept originally introduced in the context of meteorology, where chaotic behavior is found in deterministic nonlinear systems. An individual's behavior can affect others and cause nonlinear reactions in society as a whole. Phenomena such as the spread of rumors or social

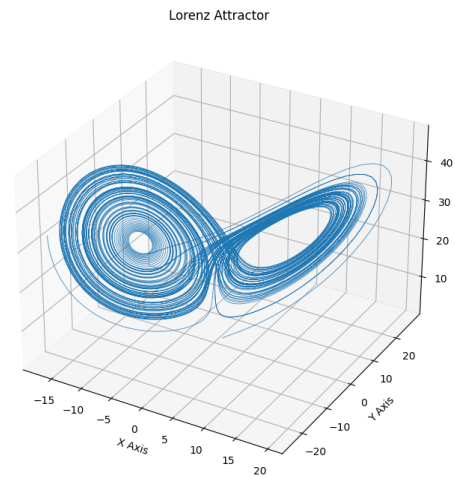

Fig. 6: Case Study:Lorenz Attractor

panic are sensitive to initial conditions, and small triggers can produce large-scale effects.

### 1.3.1 Mathematical Concepts Lorenz Attractor Poincaré Section

Define the parameters for the Lorenz system(Case study):

$$\sigma = 10.0, \quad \beta = \frac{8}{3}, \quad \rho = 28$$

The Lorenz system of differential equations is given by:

$$\frac{dx}{dt} = \sigma(y - x),$$
$$\frac{dy}{dt} = x(\rho - z) - y,$$
$$\frac{dz}{dt} = xy - \beta z,$$

where $x$, $y$, and $z$ make up the state vector of the system.

With some applications in traffic flow research, the application of the concept of Poincaré cutting planes in these social phenomena is important because vehicle flows on roads and traffic networks can cause unpredictable traffic jams as local increases or decreases in density create nonlinear feedbacks, which can lead to unpredictable system s cyclical elements and stability. For example, in economic markets, analyzing the behavior of stock prices during a particular period may help us better understand market trends. In the spread of social panics and rumors, a clearer picture of the patterns of spread may be obtained by capturing the disconnect between the way information propagates at different points in time.

It can also be suggested in terms of patterns and periodicity. Patterns such as those found in the Poincaré cutting plane of the Lorenz attractor are also found in social systems. Examples include economic cycles, political cycles, and fashion cycles. While these cycles may exhibit predictable pat-terns, they contain uncertainties that cannot be completely predicted.

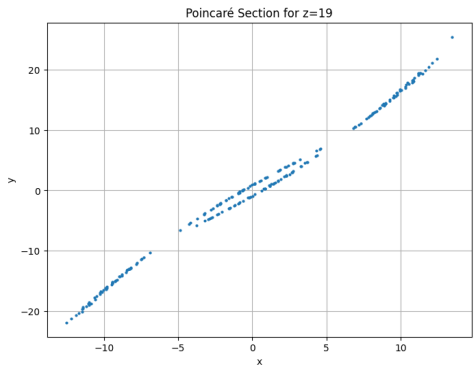

Fig. 7: Case Study:Poincaré Section for $z$=19, from Lorenz Attractor

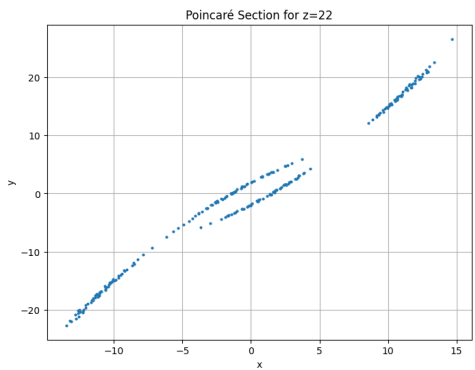

Fig. 8: Case Study:Poincaré Section for $z$=22, from Lorenz Attractor

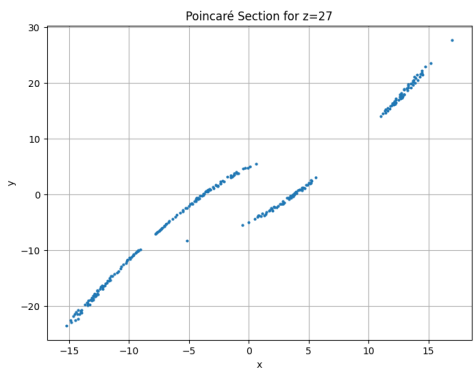

Fig. 9: Case Study:Poincaré Section for $z$=27, from Lorenz Attractor

Particular attention should be paid to the importance of initial conditions. The idea in chaos theory that subtle differences in initial conditions can lead to large differences in outcomes is similar in social phenomena. Slight differences in political, economic, and cultural conditions can drastically alter the course of history and social trends.

System like the Lorenz attractor, where system constraints and degrees of freedom are preserved, is characterized by parameters that impose constraints on its behavior. In society and in speech, factors such as laws, norms, and cultural values limit the range of behavior, but still allow for a wide variety of behaviors. The complexity of interactions is also important and the Poincaré cutting plane of the Lorenz attractor shows how interactions within a system can cause complex dynamic behavior. In society, too, interactions among individuals, groups, and organizations produce unexpected results and hold promise for model development into interpretable forms.

Figure 7, Figure 8, Figure 9 shows, The Poincaré cross section shown above is an analytical technique for understanding periodic or chaotic behavior in the state space of a dynamical system. Poincaré sections in attractor and range attractor structures are used to observe the complex motion of a 3-dimensional system in a 2-dimensional section by plotting the points at which the trajectory of the system passes through the section under specific conditions (in this case, specific $z$ values).

### 1.3.2 (1) For $z$=19

The points are arranged linearly, which suggests that the system has periodic behavior. Each point is part of a periodic pattern where the system returns to the same state. This linear arrangement may reflect a limit cycle (closed orbit). Limit cycles are solutions that exhibit stable periodic behavior under constant conditions in nonlinear dynamics.

### 1.3.3 (2) For $z$=27

Here too, the points are arranged in a constant pattern, but they are more dispersed and the distances between them are larger. This indicates that the system is behaving in a more complex or highly periodic manner. At higher z-values, it is common that the behavior of the system tends to be more complex or chaotic. This graph shows that the system may take on several different states, suggesting a transition to chaotic behavior. These cross sections help us understand how the behavior of the system changes with the value of $z$. However, these graphs do not directly indicate that a bifurcation is occurring. In order to clearly indicate a bifurcation, it would be necessary to track in detail how the behavior of the system changes as the value of z is gradually changed. To explore bi-furcation in the Poincaré cross section data, one would need to compare cross sections at different parameter values and look for quantitative changes, such as transition points from stable to unstable behavior or from periodic to chaotic behavior, and from these trends and model computational interpretations, various complex The understanding of these trends and their interpretation in terms of model calculations will also be considered.

## 1.4 Complex Networks

Drawing on insights gained from Strogatz's 2001 exploration of complex networks, the book delves into the architecture of virtual communities and reveals how online interactions are often on the knife edge of change. The sudden rise and fall of trends and the rise and fall of collective attention reflect the dynamics of saddle points and nodal bifurcations, where rising tensions within a network can trigger sudden changes and bring a once niche topic to global attention. In this digital age, where the communication landscape is constantly changing, we investigate the nonlinear dynamics at work in online interactions. We explore how subtle swings in individual behavior can lead to massive shifts in public opinion. This is reminiscent of the supercritical bifurcations described by Strogatz, where the stability of a system changes and creates new forms of order. In the maze of digital interconnectivity, the behavior of online communities emerges as a maze-like puzzle, each piece formed by the complex interplay of individual behavior and group dynamics. The digital environment is fertile ground for certain spontaneous orders and emergent phenomena that have long intrigued scientists and sociologists. Here, in the vastness of the Internet, the principles of nonlinear dynamics and bifurcation theory would be given new life and repurposed to decipher the enigmatic patterns of online engagement. Our investigation is an adventure into the heart of digital communication, aiming to chart the invisible currents that determine the flow of discourse in virtual communities. By integrating case studies and numerical simulations, we aim not only to elucidate the current state of online engagement, but also to capture the delicate balance between the spontaneous order of social harmony and its disruptive potential, and to build a model (digital chaos) to predict future flows of online engagement. We aim to From the viral spread of memes to rallies around social movements, the narrative of virtual communities is punctuated by such bifurcations. Pitchfork bifurcation, a phenomenon articulated by Strogatz, serves as a metaphor for the moment when a single online discourse diverges into divergent streams of thought. The saddle node bifurcation shows similarities to the tipping points of online campaigns that suddenly gain or lose momentum, reflecting the complex network structure detailed in Strogatz's later review (2001). Strogatz delves into the architecture of virtual communities, revealing how online interactions are often on the knife-edge of change. The sudden rise and fall of trends and the rise and fall of collective attention reflect the dynamics of saddle points and nodal bifurcations, where rising tensions within a network can trigger sudden changes, and where once niche topics can gain global attention. The rich theoretical landscape painted by the work of Guckenheimer and Holmes (1983) and further informed by the applied bifurcation exploration of Kuznetsov (2004) is a rich and rich theoretical landscape. Their rigorous analysis provides a mathematical foundation to apply to social sciences that seek to unravel the tapestry of online behavior through the lens of hypercritical bifurcations. In this context, these bifurcations represent the point at which the dominant narrative within a community becomes unstable and gives way to a new and emerging narrative that redefines the collective consciousness.

## 1.5 Blend of theory, Complex Dynamics of Online Behavior

This research is a blend of theory and practice, aimed at unraveling the complex dynamics of online behavior. We propose to integrate ideas from the rich tradition of dynamical systems and the emerging field of computational social science to craft an integrated understanding of how individual behaviors merge with powerful currents of collective action to shape our digital existence.

## 1.6 Stability of Homoclinic, Heteroclinic Orbits

In the summary of this paper, we will also touch on future research issues. In recent years, the theory of nonlinear dynamical systems has become an indispensable tool for the analysis of complex systems exhibiting unpredictable behavior in the natural and social sciences. In particular, Melnikov's method has been established as a fundamental technique for evaluating the stability of homoclinic and heteroclinic orbits in dynamical systems and for understanding the emergence of chaos. This has important implications for understanding how a system can transition to chaotic motion due to minute perturbations. In this study, we use this mathematical framework to examine specific applications in various fields.

## 1.7 Homoclinic Orbit in a Pendulum System

The pendulum's equation of motion is given by a second-order differential equation which can be expressed as a system of first-order equations:

$$\begin{cases} \frac{dy_1}{dt} = y_2, \\ \frac{dy_2}{dt} = -\sin(y_1), \end{cases}$$

where $y_1$ is the angular position and $y_2$ is the angular velocity of the pendulum.

The time span for the simulation is from 0 to 10 seconds, and the initial conditions are an angle of $2\pi$ radians (which is equivalent to 0 radians due to periodicity, suggesting the pendulum is at the equilibrium position) and an angular velocity of 100 radians per second. These are almost the initial conditions for a fixed point:

$$t_{\text{span}} = (0, 10), \quad y_0 = [2\pi, 100].$$

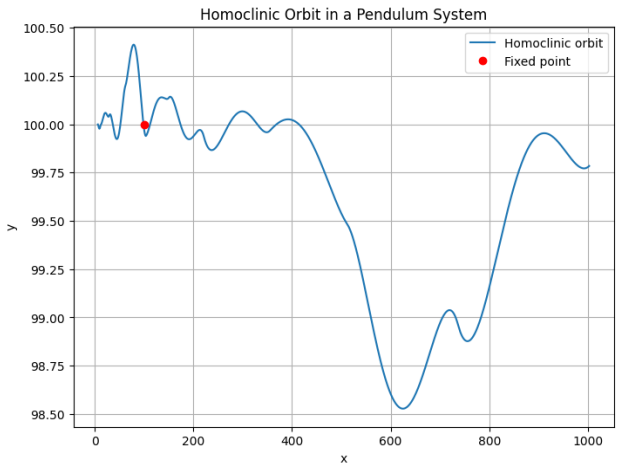

Fig. 10: Homoclinic Orbit in a Pendulum System

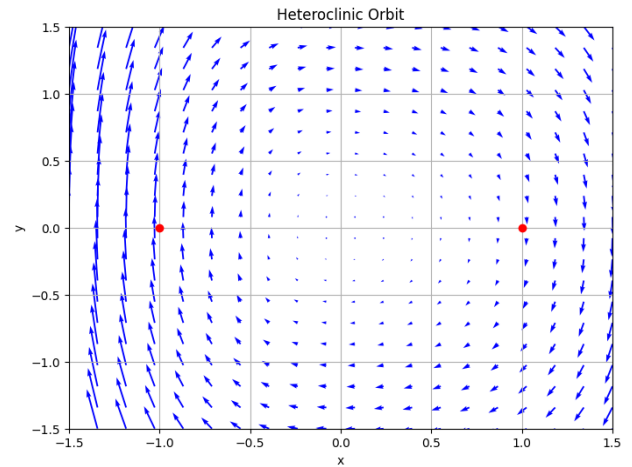

Fig. 11: Heteroclinic Orbit

## 1.8 Heteroclinic Orbits

Define the vector field function $f : \mathbb{R}^2 \to \mathbb{R}^2$ as follows:

$$f(x, y) = (y, -x + x^2 - x^3).$$

This function returns a vector field where the first component is $y$ and the second component is a cubic polynomial $-x + x^2 - x^3$ in $x$.

Create a grid for the vector field using the meshgrid function:

$$X, Y = \text{np.meshgrid}(x_{\text{values}}, y_{\text{values}})$$

Then compute the vector field $U, V$ using the defined function:

$$U, V = \text{vector\_field}(X, Y).$$

In financial markets, Farmer (1999) explored how models incorporating nonlinear dynamics can help us better understand the extreme behavior of markets. Melnikov's method, in this context, may provide new insights in predicting market abruptness. In ecosystems, on the other hand, Hastings et al. (1993) revealed the chaotic properties inherent in ecosystem dynamics and provided important implications for their management and conservation. In engineering, Moon (1987) applied Melnikov's method to the stability analysis of mechanical systems and pioneered a new understanding of vibration problems.

Furthermore, this study attempts to extend these scientific findings to applications in the context of the social sciences. Social phenomena often exhibit nonlinearities and small events can cause large changes. Melnikov's methods could be used to model social transitions such as political shifts, economic shocks, and shifts in cultural paradigms. This study aims to explore how Melnikov's method can also provide theoretical insights and practical tools in the social sciences.

This introduction clarifies the purpose and scope of the study by setting the multifaceted application of Melnikov's method as a backdrop for the study and presenting the possibilities and challenges in applying it to real-world problems.

# 2. Preview Research

## 2.1 Bifurcation Theory

In this paper, we aim to develop a research approach that correlates three types of bifurcation theories with social phenomena. First, let's talk about saddle node bifurcations. This theory is usually discussed in the context of natural sciences such as physics and biology, but is less directly common in social science contexts. Still, there are attempts to apply the saddle node bifurcation concept to social science, particularly to modeling social change and crises. For example, bifurcation theory is sometimes used to explain rapid changes in socio-economic systems or political changes when they reach a certain critical point.

Strogatz, S. H. (1994) was a foundational text on nonlinear dynamics, providing an introduction and description of various bifurcations, including pitchfork bifurcations. Strogatz, S. H. (2001) explores the structure and dynamics of complex networks, and this review paper is one of his most important contributions to the field of network theory. In Strogatz, S. H. (2003), "Sync: The Emerging Science of Spontaneous Order."-an extensive exploration of the synchronous phenomena found in nature and human society. Watts, D. J., and Strogatz, S. H. (1998). In "Small World Networks" - This paper, co-authored by Strogatz and Duncan Watts, introduces small-world networks and their properties and is a landmark study in the field of network theory. Strogatz, S. H. (2005), this paper on time crystals is a study of the nonlinear dynamics of systems with temporal periodicity. Guckenheimer, J., and Holmes, P. (1983) provided a deep understanding of the bifurcation theory of dynamical systems and described a theoretical framework for pitchfork bifurcations. Kuznetsov, Y. A. (2004) is a comprehensive reference on applied bifurcation theory and delves into the mathematical details of pitchfork bifurcations. Wiggins, S. (2003) is an introduction to bifurcation theory and chaos theory and deals with the concept of pitchfork bifurcations. Marsden, J. E., and McCracken, M. (1976) focuses on hop bifurcations, but also touches on related topics, including pitchfork bifurcations. Seydel, R. (2010) provided a practical approach to bifurcation theory, explaining various types of bifurcations, including transcritical bifurcations. Guckenheimer, J., and Holmes, P. (1983) deals with the detailed theory of dynamical systems and their bifurcations and provides a basic understanding of transcritical bifurcations.Kuznetsov, Y. A. (2004) focuses on applied bifurcation theory and delves deeply into the mathematical background of transcritical bifurcations.Wiggins, S. (2003) is an introduction to nonlinear dynamical systems and chaos and includes a section on transcritical bifurcations. Strogatz, S. H. (1994) is a well-known text as an overview of nonlinear dynamics and contains an accessible explanation for understanding transcritical bifurcations. Marsden, J. E., and McCracken, M. (1976) provides a detailed discussion of hop bifurcations, but also touches on other branching theories, especially transcritical bifurcations. Strogatz, S. H. (1994)This text provides a comprehensive treatment of nonlinear dynamics and an accessible account of saddle-node bifurcations. Kuznetsov, Y. A. (2004) dedicated to applied bifurcation theory, this book provides a detailed study of various bifurcation phenomena, including saddle-node bifurcations. Guckenheimer, J., and Holmes, P. (1983). delves into the theoretical aspects of saddle node bifurcations, covering both the fundamentals and applications of dynamical systems. The intersection of sociophysics and digital communication has opened a novel vista for understanding human behavior and opinion dynamics. The digital age has transformed how opinions are formed, evolved, and propagated, especially on platforms such as social networking services (SNS). Capturing this complexity requires innovative models that reflect both the immediacy of information exchange and the myriad cognitive processes at play within individuals. This study introduces a comprehensive numerical simulation model that encapsulates these aspects using a phase field approach to represent the understanding and cognition of information as continuous variables.

## 2.2 Case Study:Bifurcation

Pitchfork bifurcation is a phenomenon studied in dynamical systems theory, particularly in the field of nonlinear dynamics, and refers to the point at which the behavior of a system changes qualitatively in response to changes in parameters. Pitchfork bifurcation, in dynamic systems theory, refers to a phenomenon in which a system bifurcates from one stable point to three stable points in response to changes in parameters. When the concept of pitchfork bifurcation is applied in a social science context, it is often used as a model for understanding points of change in society, especially political, economic, and cultural fluctuations. However, it is difficult to link it directly to concrete social science literature and is often used as a metaphor or applied as a theoretical framework to explain complex social phenomena. A pitchfork bifurcation is characterized by the transition of a dynamical system from a stable fixed point to an unstable fixed point at the bifurcation, resulting in a symmetrical pair of stable fixed points at the same time. It is known that this causes the structure of the solution of the system to change significantly at the branch point. Transcritical bifurcation is one of the phenomena observed in bifurcation theory of dynamical systems. It refers to a type of bifurcation in which two fixed points are exchanged when the parameters of the system pass a certain value, and during this process one fixed point changes from stable to unstable or vice versa. Although case studies in the social sciences about the transcritical bifurcation are difficult to find direct counterparts, there may be cases where this mathematical concept is metaphorically applied to specific situations and models in the social sciences. Transcritical bifurcation refers to a process in which a stable point becomes unstable as a certain parameter changes, and in the context of social science, it is sometimes used to express how the equilibrium state of a social system changes. As in previous research, an analogy can be drawn from the Pitchfork bifurcation society: Political change: A political system changes from a stable state (e.g., one-party dictatorship) to a small change (e.g., new policies or changes in leadership). This can be seen as a branching out into multiple possibilities (e.g. multiparty system, full democracy, anarchy). Phenomenon on social media: A certain trend or hashtag initially has one meaning, but due to a sudden incident or a celebrity's comment, it branches into many different meanings and subgroups, and begins to behave differently. The social analogy of a saddle node bifurcation is an economic crisis: when the economy reaches a certain point, a small external shock causes a stable market to suddenly collapse, resulting in a completely different economic situation, a deep recession. It looks like we are on the verge of rapid inflation. A phenomenon on social networking sites is that when a company's scandal comes to light, the company's reputation plummets due to a single critical report, and what was once a neutral social evaluation quickly turns negative.

In the social analogy of transcritical bifurcation, changing social trends: how one social custom or cultural trend is replaced by an entirely new trend due to subtle social changes or value shifts. (For example, a major change in career path due to a change in work values). Social media phenomenon: The lifestyle promoted by an influencer completely loses support from his followers due to his actions or scandals, and the opposite image or movement becomes mainstream. Although these examples are different from actual mathematical branches, they serve as analogical tools for understanding social trends and phenomena on social media. When applying mathematical models to the social sciences, such metaphors can provide a useful framework for grasping the complexity and dynamics of phenomena. However, it is assumed that actual social phenomena are much more complex than mathematical models and contain many unpredictable elements. Pitchfork bifurcation can be applied to consensus building to show the progression of consensus building: the moment a group's opinion splits and new subgroups are formed during a discussion. Under certain conditions (e.g., presentation of important information), a group's opinion shifts from homogeneity to polarization. When saddle node bifurcation is applied to conflicts of opinions, rapid changes in opinions occur: When a community maintains a delicate balance of opinions on a topic, one event can cause one opinion to completely disappear, and the opposite It can indicate the moment when an opinion becomes dominant. Applying the transcritical bifurcation to the evolution of opinions, we see opinion reversals: when a social consensus exists about an issue, new information or a change in circumstances completely reverses the social consensus, It can be expressed that an opinion that was once a minority becomes mainstream.

The mathematics of mass panic at the Pitchfork Bifurcation shows that when people sense danger, they initially act calmly, but once a certain critical point is exceeded, they suddenly panic and begin to behave in unpredictable ways. In other words, collective behavior rapidly shifts from a calm state to a panic state. Change in traffic flow: The moment when the flow changes from smooth to congested at a point where the road is congested. This change occurs when the vehicle inflow rate exceeds a critical value. In the mathematics of mass panic at the saddle node branch, a group marching in a demonstration: A demonstration march is established when there are a certain number of participants, but when the number falls below a certain number, it naturally disbands. This critical point in the number of individuals can be regarded as the saddle node bifurcation point. Mathematics of mass panic at transcritical bifurcation: Flow of shoppers: When the flow of shoppers goes from quiet to a certain level of congestion at a supermarket, etc., the movement of customers changes significantly. For example, when the number of people increases, actions such as changing the travel route are taken. These examples illustrate how people's collective behavior shifts to qualitatively different states in response to changes in external conditions. Nonlinear dynamical systems are often applied to the movement of people and the behavior of groups, and can be modeled through these bifurcation phenomena. Understanding and predicting collective behavior is important in many practical areas, such as urban planning, event management, and emergency evacuation planning.

## 2.3 Case Study:Bifurcation of Nature

An example of a natural phenomenon studied in pitchfork bifurcation is the transition of a fluid from laminar to turbulent flow: a phenomenon in which the flow suddenly changes from stable laminar flow to unstable turbulent flow when a certain critical velocity is exceeded. Magnetization of a magnetic material: When the temperature drops below a critical point, a magnetic material changes from an unmagnetized state to a magnetized state, creating two stable states with respect to the direction of the magnetic field. A research example of a natural phenomenon in saddle node bifurcation is the extinction point in an ecosystem: When the conditions under which a species can survive within an ecosystem reach a breaking point, that species can suddenly become extinct. Global climate change: Historical climate data shows that once the Earth's climate reaches a certain threshold, it will rapidly transition into glacial or interglacial periods. An example of a natural phenomenon studied in the transcritical bifurcation is the spread of a disease: When an infectious disease exceeds a threshold of infectivity, it moves from a stable state where the number of infected people is small to a new stable state where the infection spreads. Equilibrium state of a chemical reaction: A chemical reaction proceeds under certain conditions and reaches a new equilibrium state by changing the ratio of reactants and products. These examples of natural phenomena have in common with mathematical bifurcations in that systems behave differently depending on changes in certain parameters. These processes in nature are often predictable and explained by quantitative models. On the other hand, social phenomena and trends on SNS are more non-linear, and quantitative prediction is often difficult due to the complexity of human decision-making and collective behavior. Pitchfork bifurcation, saddle node bifurcation, and transcritical bifurcation are typical bifurcation phenomena observed in dynamical systems theory. These are found in a wide variety of scientific fields, including physics, biology, chemistry, economics, and engineering. Each branch represents how the system transitions to a different stable state in response to changes in parameters. Specific examples of each branch are introduced below. Pitchfork Bifurcation: Physics: When the magnetization of a magnet is used as a parameter to control temperature, the direction of the magnetization reverses after a certain critical point is exceeded.

This can be considered a pitchfork bifurcation to magnetization. In biology, it is a phenomenon in which the population of microorganisms shifts to different growth patterns as the concentration of nutrients increases or decreases. In physics, saddle node bifurcation is a phenomenon in which the current increases rapidly at a certain voltage in the relationship between voltage and current in a diode in an electronic circuit. In biology, a situation in which competition between species leads to extinction when the population of a species falls below a critical value. In physics, transcritical bifurcation is a phenomenon in fluid flow where the flow pattern shifts from laminar to turbulent due to changes in the Reynolds number (a dimensionless number that represents the characteristics of fluid flow). It can be said that the type of reaction that occurs in a chemical reactor varies depending on the concentration of the reactants. For example, we can say that if the concentration of a reaction in the presence of a catalyst changes, the reaction pathway changes. Pitchfork bifurcations, saddle node bifurcations, and transcritical bifurcations in online traffic and networks can exhibit the following phenomena: In the Pitchfork bifurcation, website traffic spikes: A news article or event triggers a sudden increase in website traffic. At first, the number of accesses increased steadily, but at a certain point the number of accesses increased explosively, and there was a possibility that the server would go down. Changes in trends on social media: After a certain hashtag or topic gains a certain level of popularity, it suddenly becomes viral and becomes established as a trend. A saddle node bifurcation is a loss of network connectivity: A particular service or node on the network continues to be stressed to a critical point, resulting in a complete loss of connectivity and communication becomes impossible.

In the transcritical bifurcation, there is a qualitative change in network traffic: As the amount of data traffic continues to increase, the state of the network changes, and the delay is small at first, but after a certain point, the delay becomes normal. These phenomena are important factors to be considered when analyzing online information propagation and traffic flow dynamics. For example, when a large amount of access is expected, precautions such as increasing the website's scaling ability and server resource allocation may be necessary. Additionally, understanding these bifurcation phenomena can help you assess network vulnerabilities and develop appropriate load-balancing strategies.

In summary, our study presents a novel approach to modeling opinion dynamics in digital communication, offering a theoretical and computational framework for analyzing the complex interplay of individual cognitive processes and collective behaviors. Through this work, we strive to provide insights into the digital citizenry's opinion formation and evolution, contributing to the broader field of sociophysics and its applications in understanding the digital society.

# 3. Bifurcation Theories to Social Phenomena

In this paper, we aim to take a research approach that maps three types of bifurcation theories to social phenomena. The first one is the saddle node bifurcation. This theory is usually discussed in the context of the natural sciences, such as physics and biology, but is less common directly in the context of the social sciences. Nevertheless, there have been some attempts to apply the concept of saddle node bifurcation to the social sciences, particularly in the modeling of social changes and crises. For example, bifurcation theory is sometimes used to explain abrupt changes in socioeconomic systems or political changes when certain critical points are reached.

Pitchfork bifurcation is a phenomenon studied in dynamical systems theory, especially in the field of nonlinear dynamics, and refers to the point at which the behavior of a system changes qualitatively in response to changes in its parameters. In dynamic systems theory, pitchfork bifurcation refers to the phenomenon in which a system bifurcates from one stable point to three stable points in response to changes in parameters. When the concept of pitchfork bifurcation is applied in the context of the social sciences, it is often used as a model for understanding social change points, particularly political, economic, and cultural fluctuations. However, it is difficult to connect this directly to the specific social science literature and is often used as a metaphor or applied as a theoretical framework to explain complex social phenomena. Pitchfork bifurcation is characterized by the fact that the dynamical system goes from a stable fixed point to an unstable fixed point at the bifurcation point, resulting in a symmetric pair of stable fixed points at the same time. This is known to cause a significant change in the structure of the solution of the system around the bifurcation point. Transcritical bifurcation is one of the phenomena found in the bifurcation theory of dynamical systems. It refers to a type of bifurcation in which two fixed points are exchanged when the parameters of the system pass a certain value, and during this process one fixed point changes from stable to unstable or vice versa. Case studies in the social sciences on transcritical bifurcation are difficult to find direct counterparts, but there may be cases where this mathematical concept is applied metaphorically to specific situations or models in the social sciences. Transcritical bifurcation refers to the process by which a stable point becomes unstable as some parameter changes, and in the context of the social sciences it is sometimes used to describe the changing equilibrium state of a social system.

### 3.1 Opinion Dynamics to Saddle Node Bifurcation

#### 3.1.1 Modeling for Saddle Node Bifurcation

The dynamical system is defined by the steady state equation:

$$f(\text{Opinion\_A}, k_{AB}, k_{AC}, a_{\text{media}}, b_{\text{media}}) =$$
$$k_{AB} \cdot \text{Opinion\_A}+$$
$$k_{AC} \cdot \text{Opinion\_A}^2+$$
$$a_{\text{media}} + b_{\text{media}}$$

where:

Opinion_A is the opinion level of Cluster A which we are solving for in the steady state.

$k_{AB}$ is a parameter representing the interaction coefficient between Cluster A and B.

$k_{AC}$ is a parameter representing the interaction coefficient within Cluster A itself, initially set to 0.05.

$a_{\text{media}}$ is a parameter representing media influence towards a positive opinion, initially set to 100.

$b_{\text{media}}$ is a parameter representing media influence towards a negative opinion, initially set to $-100$.

The function `search_bifurcation` iteratively computes the steady state Opinion_A for a range of $k_{AB}$ values by finding the roots of the steady state equation.

The bifurcation diagram is plotted with $k_{AB}$ on the x-axis and the steady state values of Opinion_A on the y-axis. The diagram is used to visualize the points where the behavior of the system changes qualitatively, known as bifurcation points.

Additionally, the code computes the first and second derivatives of the steady states with respect to $k_{AB}$ to find the inflection points where the curvature changes. These points are identified where the second derivative is closest to zero, indicating a change in the concavity of the bifurcation curve.

The fixed points and the inflection point coordinates are then outputted, with the latter being particularly significant in the study of dynamical systems as they can signal a change in stability or the presence of a bifurcation.

#### 3.1.2 Discussion of Saddle Node Bifurcation

The provided graph appears to be a saddle-node bifurcation diagram, which is a type of bifurcation where two equilibrium points (one stable, one unstable) collide and annihilate each other as a parameter changes. The bifurcation diagram shows how the steady state of $Opinion_A$ varies with the parameter $k_{AB}$. From the inflection point coordinates $(k_{AB}, Opinion_A) = (0.10687, 2.14812 \times 10^{-14})$, we can infer that the steady state of $Opinion_A$ is essentially zero when $k_{AB}$ is at approximately 0.107. This value is so close to zero that, for all practical purposes within the model, it can be considered as zero. This suggests that at this critical value of $k_{AB}$, a bifurcation occurs: $Opinion_A$ transitions from a non-zero steady state to zero. The critical value of $k_{AB}$ where this transition occurs is significant because it indicates the parameter's threshold beyond which the opinion can no longer sustain itself in the population. The exact nature of the transition—whether it is a gradual decay or a sudden collapse—cannot be determined solely from the inflection point, but typically in saddle-node bifurcations, the transition is abrupt.

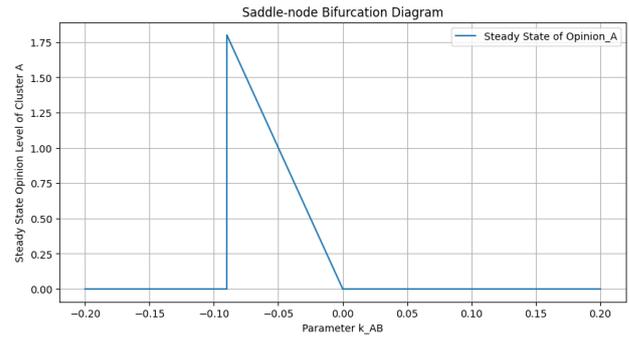

Fig. 12: Saddle-node Bifurcation Diagram

In terms of social dynamics, this model might represent a scenario where increasing the influence of counter-opinions or external factors (represented by the parameter $k_{AB}$) gradually weakens a previously stable opinion until it can no longer maintain itself and suddenly disappears. This could apply to various social systems where opinions, behaviors, or beliefs can change dramatically when influenced by external factors beyond a critical point.

The graph provided appears to show a saddle-node bifurcation diagram for a system that models the steady-state level of an opinion cluster Opinion_A as a function of the parameter $k_{AB}$. In the context of dynamical systems, a saddle-node bifurcation is where two fixed points of the system (a stable and an unstable point) collide and annihilate each other as a parameter is varied.

From the function you've provided:

$$f(\text{Opinion\_A}, k_{AB}, k_{AC}, a_{\text{media}}, b_{\text{media}}) =$$
$$k_{AB} \cdot \text{Opinion\_A}+$$
$$k_{AC} \cdot \text{Opinion\_A}^2+$$
$$a_{\text{media}} + b_{\text{media}}$$

It appears that $k_{AB}$ and $k_{AC}$ are parameters that influence the opinion level directly and quadratically, while $a_{\text{media}}$ and $b_{\text{media}}$ are constants that could represent media influence or bias.

The graph suggests that as $k_{AB}$ increases past a critical point (near 0 on the x-axis), the steady-state opinion level abruptly decreases to zero. This could indicate a scenario in which increasing the influence of a particular factor or opinion (represented by $k_{AB}$) leads to a sudden change in the public consensus or the prevalent opinion in a community.

In social terms, this could be interpreted as follows:

**Pre-Bifurcation:** When $k_{AB}$ is negative and large in magnitude, there's a high, steady opinion level, suggesting strong agreement or a dominant narrative within a population.

**At Bifurcation Point:** As $k_{AB}$ approaches the critical value from the left, it signifies a delicate balance where a small change in the parameter could result in a significant shift in the public opinion.

**Post-Bifurcation:** Once $k_{AB}$ passes the critical value (to the right of the bifurcation point), the opinion level drops to zero, indicating a collapse or sudden shift in the dominant opinion or a consensus point, which could be due to a loss of credibility or a shift in public sentiment.

This kind of model and analysis is relevant in understanding how public opinion can be stable under certain conditions but can suddenly shift when critical parameters, possibly representing external influences or internal dynamics within the population, reach certain thresholds. It can be applied to various contexts, such as political sentiment, market dynamics, or cultural trends.

Saddle node bifurcation (or limit point bifurcation) is a phenomenon in nonlinear dynamics in which stable and unstable fixed points merge and disappear. Such bifurcations occur when the parameters of the system exceed critical values, often resulting in abrupt changes in the behavior of the system. The results can be considered in the context of opinion dynamics from the following perspectives.

### (1) Consideration as a social phenomenon

Opinion formation in society is an example of nonlinear dynamics. In the context of saddle node bifurcation, the equilibrium state of opinions within a social group may disappear due to changes in certain parameters (e.g., social pressure or information flow). This suggests a phenomenon in which an argument can exist stably under a certain pressure, but rapidly lose support when the pressure changes. For example, this might include a "taboo" moment when a particular opinion suddenly becomes unacceptable to society.

### (2) Consideration of media as an influence

The media has a significant impact on opinion formation in society, and the amedia and bmedia parameters model media influence as both positive and negative. This indicates that the media provides both information supporting (positive influence) and opposing (negative influence) an opinion. When a saddle node bifurcation occurs, a sudden change in media influence can result in a significant shift in societal opinion. For example, this is the case where society's opinion changes as a result of media coverage that is biased toward one side of an issue.

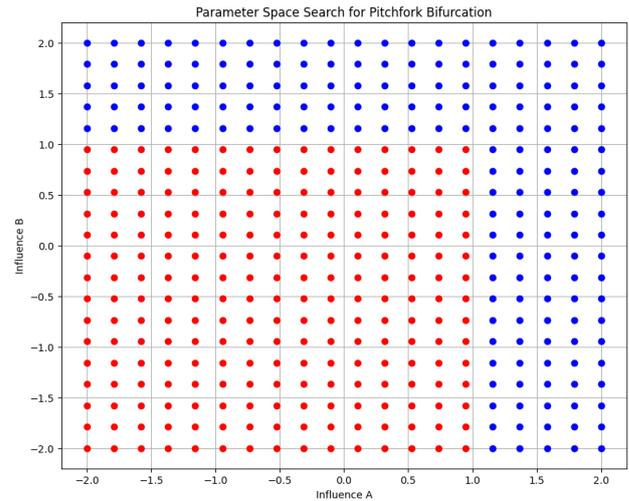

Fig. 13: Parameter Space Search for Pitchfork Bifurcation

### (3) Consideration as consensus building

Consensus building is an important aspect of opinion dynamics, and the saddle node bifurcation suggests an important turning point in the consensus building process. Once the bifurcation point is crossed, the stable consensus of opinion that previously existed can no longer be achieved. This phenomenon refers to a situation in which a social group has formed a consensus on an opinion, but new information or changes in external pressures cause the consensus to collapse and a new equilibrium of opinion becomes necessary. For example, a change in social values or the rise of a new generation may require a new discussion of previously agreed upon issues.

When considering the characteristics of saddle node bifurcation, it is important to note that these social, media, and consensus-building processes can change significantly with only subtle changes in the parameters of the system. This indicates that social opinion is very complex and can vary greatly with even subtle changes.

## 3.2 Opinion Dynamics to Pitchfork Bifurcation

### 3.2.1 Modeling for Pitchfork Bifurcation

We consider a system of four interacting opinions modeled by the following differential equations:

$$\frac{d\text{Opi}_A}{dt} = -\text{Opi}_A + \text{influence}_A \cdot \text{Opi}_A \cdot (1 - \text{Opi}_A^2), \quad (1)$$

$$\frac{d\text{Opi}_B}{dt} = -\text{Opi}_B + \text{influence}_B \cdot \text{Opi}_B \cdot (1 - \text{Opi}_B^2), \quad (2)$$

$$\frac{d\text{Opi}_C}{dt} = -\text{Opi}_C + \text{influence}_C \cdot \text{Opi}_C \cdot (1 - \text{Opi}_C^2), \quad (3)$$

$$\frac{d\text{Opi}_D}{dt} = -\text{Opi}_D + \text{influence}_D \cdot \text{Opi}_D \cdot (1 - \text{Opi}_D^2). \quad (4)$$

The parameters $\text{influence}_A$, $\text{influence}_B$, $\text{influence}_C$, and $\text{influence}_D$ represent the respective influences on opinions A, B, C, and D. For the bifurcation analysis, we consider $\text{influence}_C = \text{influence}_D = 1.0$ as fixed parameters and vary $\text{influence}_A$ and $\text{influence}_B$.

The initial conditions for each opinion state are set to 0.1. The time grid for the simulation spans from 0 to 10 at 100 evenly spaced intervals.

A parameter sweep is performed for $\text{influence}_A$ and $\text{influence}_B$ to identify conditions that may lead to a pitchfork bifurcation. The criteria for identifying a potential bifurcation is that both steady states of Opinion A and Opinion B are near zero (within a threshold of 0.1).

A bifurcation diagram is plotted for Opinion A by varying $\text{influence}_A$ and observing the resulting steady states. The steady state of Opinion A is determined by the last value of the state after integration over the time grid.

### 3.2.2 Discussion of Pitchfork Bifurcation

We analyze a dynamical system with a single state variable, Opinion A, which evolves over time according to the influence of a single parameter. The system is modeled by the following differential equation representing a typical pitchfork bifurcation:

$$\frac{d\text{Opinion}_A}{dt} = \text{influence}_A \cdot \text{Opinion}_A - \text{Opinion}_A^3 \quad (5)$$

Here, $\text{Opinion}_A$ is the state variable and $\text{influence}_A$ is the bifurcation parameter. The initial condition for Opinion A is set to 0.1, and we observe the system's behavior over the time interval from 0 to 10.

To construct the bifurcation diagram, we sweep the $\text{influence}_A$ parameter through a range from $-2$ to 2. At each value of $\text{influence}_A$, we integrate the differential equation over the defined time interval and record the last value as the steady state of Opinion A. This steady state is plotted against the bifurcation parameter to visualize the bifurcation diagram.

The bifurcation point, where the behavior of the system changes, is identified at $\text{influence}_A = 0$. At this point, the system transitions from a single stable steady state to multiple steady states as $\text{influence}_A$ increases past zero. The steady state of Opinion A at the bifurcation point is numerically calculated and reported.

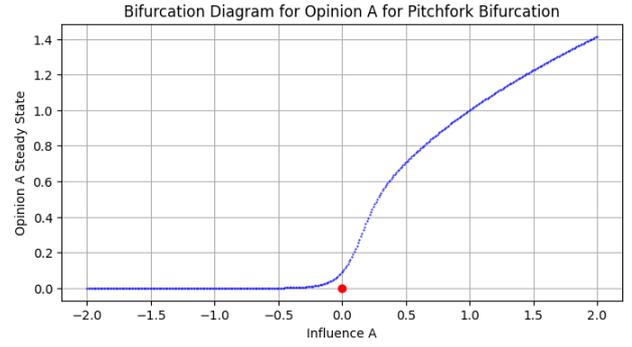

Fig. 14: For Opinion A for Pitchfork Bifurcation Steady state for Opinion A at the bifurcation point is 0.091287

This results suggests that the time rate of change of "Opinion A" is proportional to the current state of "Opinion A" scaled by "influence A" and reduced by the cube of "Opinion A". This is a simplified model results of a real-world system where opinions can change over time under the influence of external factors and self-reinforcement with a non-linear feedback.

Regarding your request for an explanation and consideration from three different perspectives:

(1) **Social Phenomena Perspective**

The pitchfork bifurcation indicates a scenario where a small change in the influence parameter can lead to a sudden and large change in the steady-state of public opinion, representing how a critical mass or tipping point can result in significant societal shifts.

Before the critical influence level (bifurcation point), there is only one stable state of opinion. Past this point, the system has the potential to adopt one of two new stable states, indicating the possibility of polarized opinions.

(2) **Media Influence Perspective**

The role of media can be seen as the parameter that can push the system past the bifurcation point, leading to the formation of strong opinions.

The media's influence can be seen in the transition from a single steady-state opinion to multiple, possibly extreme, opinions. The media's role in amplifying certain narratives can cause society to adopt a polarized view.

(3) **Consensus Formation Perspective**

The pitchfork bifurcation shows that consensus can be fragile. As influence increases, the consensus (the stable state at Opinion A = 0) can break down, leading to the formation of multiple distinct groups.

The bifurcation diagram shows that beyond a certain threshold, consensus may not be possible, and the society might split into groups with strongly held, divergent views.

The model is highly idealized, and real-world scenarios are affected by a multitude of factors, including network effects, individual biases, historical context, and more. However, this model provides a framework to understand how collective opinion can undergo sudden changes due to gradual changes in underlying parameters.

## 3.3 Opinion Dynamics to the Transcritical Branch

### 3.3.1 Modeling for Transcritical Branch

We consider a stochastic dynamical system with five clusters (A, B, C, D, E). The dynamics of the opinions in these clusters are influenced by both media effects and stochastic events represented as a system of stochastic differential equations (SDEs):

$$dO_A = \left( \frac{(\alpha_{\text{media,A}} - \beta_{\text{media,A}})}{\text{size}_A} O_A - \text{message}_A \right) dt$$
$$+ \sigma_A dW_A, \qquad (6)$$

$$dO_B = \left( \frac{(\alpha_{\text{media,B}} - \beta_{\text{media,B}})}{\text{size}_B} O_B + \text{message}_A - \text{message}_B \right) dt$$
$$+ \sigma_B dW_B, \qquad (7)$$

$$dO_C = \left( \frac{(\alpha_{\text{media,C}} - \beta_{\text{media,C}})}{\text{size}_C} O_C + \text{message}_C \right) dt$$
$$+ \sigma_C dW_C, \qquad (8)$$

$$dO_D = \left( \frac{(\alpha_{\text{media,D}} - \beta_{\text{media,D}})}{\text{size}_D} O_D - \text{message}_C + \text{message}_D \right) dt$$
$$+ \sigma_D dW_D, \qquad (9)$$

$$dO_E = \left( -\text{message}_E \frac{O_E}{\text{size}_E} \right) dt$$
$$+ \sigma_E dW_E. \qquad (10)$$

Each SDE of the form

$$dO_i = a_i(t, O_i) \, dt + b_i(t, O_i) \, dW_i \qquad (11)$$

consists of two parts: the drift term $a_i(t, O_i)$ and the diffusion term $b_i(t, O_i)$. The drift term describes the deterministic trend of the opinion state $O_i$ over time, while the diffusion term accounts for the random fluctuations, with $dW_i$ representing the increments of a Wiener process, which models the stochastic noise.

In our specific system, the drift terms for the opinion states are given by:

$dO_A$: The rate of change of opinion A is proportional to the current opinion $O_A$, adjusted by the net media influence (positive influence $\alpha_{\text{media,A}}$ minus negative influence $\beta_{\text{media,A}}$), normalized by the size of cluster A, and reduced by the number of messages $\text{message}_A$ directed at cluster A.

$dO_B$: Similarly, the rate of change of opinion B is influenced by its own media impact and size, with additional terms accounting for messages received from cluster A and sent to cluster B.

$dO_C$ and $dO_D$: These follow the same structure as above, with their respective media influences and message exchanges.

$dO_E$: The rate of change of opinion E is negatively influenced by the number of random messages $\text{message}_E$, which is a function generating random values, representing the unpredictable influence on opinion E, again normalized by the size of cluster E.

The diffusion terms for all opinions are assumed to be the same:

$b_i(t, O_i) = \sigma_i$, where $\sigma_i$ is a small constant, suggesting that the random fluctuations have the same intensity across all clusters.

The initial conditions for all opinion states are set to a small value, and the system is observed over a time interval from 0 to 100, discretized into 10,000 time steps.

where $O_i$ represents the opinion in cluster $i$, $\text{message}_i$ represents the number of messages influencing the cluster $i$, $\alpha_{\text{media,}i}$ and $\beta_{\text{media,}i}$ represent positive and negative media influences on cluster $i$, respectively. The sizes of the clusters are denoted by $\text{size}_i$, and $\sigma_i$ represents the intensity of the stochastic influence on cluster $i$ with $dW_i$ being the standard Wiener process.

The parameters $\alpha_{\text{media,}i}, \beta_{\text{media,}i}$, and $\text{message}_i$ for clusters A, B, C, and D are constants, while for cluster E, $\text{message}_E$ is a random variable representing random messages, simulated as a random integer between 100 and 10000.

### 3.3.2 Discussion of Transcritical Branch

### 1. Social Phenomenon Perspective

This model can be seen as a representation of how opinions (denoted by $O_X$) within social clusters are influenced over time. Each cluster's opinion dynamics are influenced by media parameters and internal messages. A transcritical bifurcation in this context could indicate a critical point where the system's behavior qualitatively changes, such as a shift from a consensus to a polarized state or vice versa.

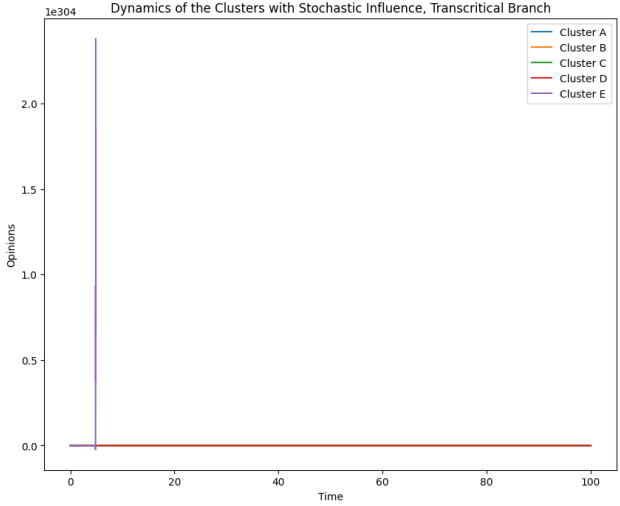

Fig. 15: Dynamics of the Clusters with Stochastic Influence, Transcritical Branch

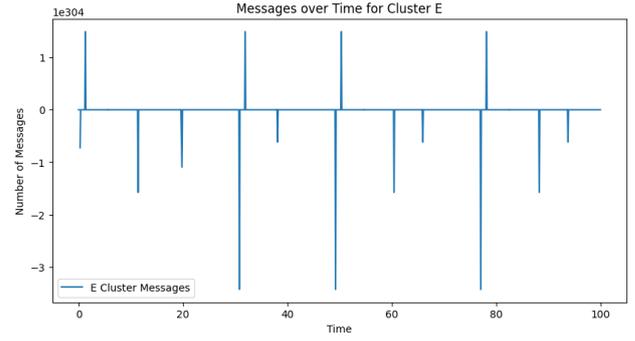

Fig. 16: Dynamics of the Clusters with Stochastic Influence, Transcritical Branch part1

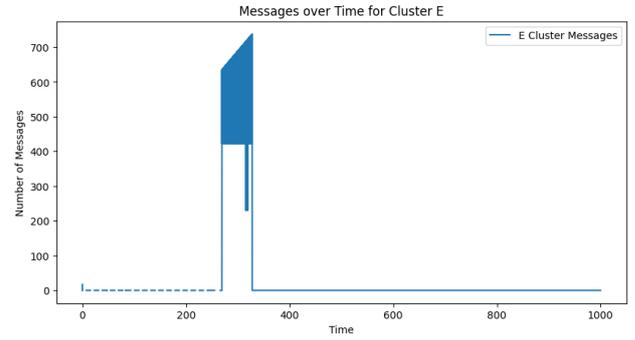

Fig. 17: Dynamics of the Clusters with Stochastic Influence, Transcritical Branch part2

## 2. Media Influence Perspective

The media's impact on opinion dynamics is modeled by parameters $\alpha$ and $\beta$, which could represent the rate of positive and negative media influence on the clusters' opinions, respectively. A transcritical bifurcation might occur when the media influence balances out the internal messages, potentially leading to a shift in the dominant opinion or the stability of the opinion distribution.

## 3. Consensus Formation Perspective

The differential equations suggest an exchange of messages between different clusters, potentially representing the interplay of opinions and how they may converge or diverge over time, leading to consensus formation or fragmentation. A transcritical bifurcation might represent a point where the system transitions from a state of consensus to one of disagreement, or it could indicate the level of media influence required to shift the public opinion from one stable state to another.

For a more detailed analysis of the transcritical bifurcation and its implications in this context, the equations would typically need to be solved analytically or numerically to observe the system's behavior near the bifurcation point.

## 3.4 Stability Analysis for Cluster E

In our dynamical system, the stability of the 'E' cluster at each time step is evaluated based on the results obtained from the simulation. For each value of the state variable $O_E$ extracted from the results, we calculate the fixed point fixed_point_E by normalizing $O_E$ with the size of the cluster size$_E$:

$$\text{fixed\_point\_E} = \frac{O_E}{\text{size}_E} \tag{12}$$

The stability condition is then assessed by comparing the negative reciprocal of the size of the cluster size$_E$ with zero:

$$\text{stability} = \begin{cases} \text{'Stable'} & \text{if } -\frac{1.0}{\text{size}_E} < 0, \\ \text{'Unstable'} & \text{otherwise.} \end{cases} \tag{13}$$

Since $-\frac{1.0}{\text{size}_E}$ is always negative due to the size being positive, this condition will always evaluate to 'Stable' for any positive size of the cluster. This indicates that the analysis in the code is not meaningful for stability determination and might require a revision to incorporate a valid stability criterion.

The two graphs you've provided show "Messages over Time for Cluster E" with different scales and possibly different time frames or data aggregation levels. In both graphs, we observe spikes in the number of messages at certain time intervals. The first graph indicates a much higher frequency and magnitude of messages than the second.

## 1. Social Phenomenon Perspective

The spikes in message frequency could reflect moments of heightened social activity or response to events. A transcritical bifurcation in social dynamics might be indicated by a sudden change in the number or sentiment of messages, suggesting a shift in public opinion or social norms. This could correspond to critical thresholds in societal factors that, when crossed, result in a significant change in social behavior or consensus.

## 2. Media Influence Perspective

The graph could be illustrating the impact of media on the cluster, with spikes representing times of intense media activity or viral events. A transcritical bifurcation here could indicate the level of media influence that leads to a change in the prevalent views within the cluster. For instance, the graph might be showing the tipping point at which media narratives begin to significantly alter the cluster's opinions.

## 3. Consensus Formation Perspective

The message patterns can also be interpreted as indicators of consensus formation or disruption within Cluster E. The sharp increases and decreases in message count could illustrate the points at which consensus is either being built or broken down. A transcritical bifurcation might manifest as a point where the message activity shifts from one stable state of consensus to another, potentially due to internal dynamics or external influences.

In both graphs, the transcritical bifurcation characteristics might be visible as moments where the message activity abruptly changes. However, without a clear understanding of the context of these messages and additional data on the cluster's internal and external interactions, it's challenging to make a precise analysis. The dynamics of message activity could be further elucidated with a model that captures the complex interplay of social forces, media influence, and the mechanisms of consensus formation within the cluster. This would likely involve a deeper analysis of the system's parameters and their influence on the stability and transitions of the cluster's opinion states.

## 1. Social Phenomenon Perspective

Also the sharp increases in message frequency might indicate rapid changes in social dynamics within Cluster E, such as a sudden organization of a social movement, collective response to an event, or a viral trend. A transcritical bifurcation in this context could reflect a critical threshold of social activity or sentiment that, once reached, leads to a new equilibrium state—such as a shift from a norm to a new social norm or from disengagement to active participation.

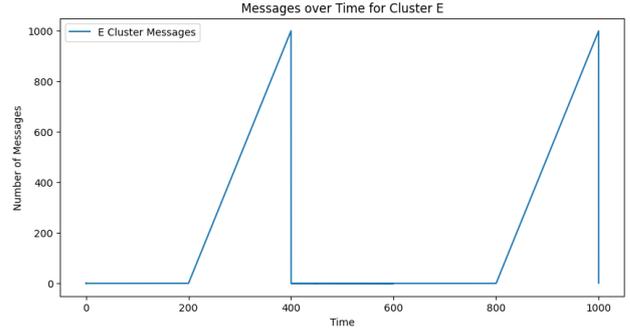

Fig. 18: Dynamics of the Clusters with Stochastic Influence, Transcritical Branch part3

## 2. Media Influence Perspective

These peaks could also signify the moments when media coverage or media-induced events impact the cluster, causing a surge in communication. The transcritical bifurcation here might represent the critical level of media impact necessary to alter the cluster's state from one of equilibrium to another—for example, from having a neutral stance to adopting a particular viewpoint influenced by the media, too.

## 3. Consensus Formation Perspective

The patterns shown in the graph can be interpreted as the process of consensus building within Cluster E. The sudden rise in messages may indicate vigorous discussion or debate leading to the formation of a consensus, while the subsequent drop could signify the resolution of discussions and the establishment of a new consensus. A transcritical bifurcation in this scenario would be the point at which the number of messages (and thus the level of active engagement) shifts dramatically, marking the transition from one consensus state to another.

# 4. Conclusion

## 4.1 Public opinion and Homo-clinic Orbits in Dynamical Systems

### 4.1.1 Modeling for Homo-clinic Orbits

We examine a two-dimensional dynamical system described by the following set of ordinary differential equations (ODEs):

$$\frac{du}{dt} = w, \tag{14}$$

$$\frac{dw}{dt} = p_1 u - \frac{u^2}{2} - \epsilon u^2 w, \tag{15}$$

where $u$ and $w$ are the state variables of the system. The parameter $p_1$ is a positive constant, and $\epsilon$ is a small perturbation parameter. The system is designed to exhibit homoclinic behavior, where the trajectory in the phase space returns to a saddle fixed point as time approaches infinity.

**Numerical Integration**

The system's stable and unstable manifolds are computed by integrating the ODEs with the initial conditions near the origin, $(0,0)$, over positive and negative time, respectively. For numerical integration, we employ the `solve_ivp` method from the SciPy library with the 'DOP853' integrator and a high tolerance for relative error, rtol = $1 \times 10^{-10}$.

The resulting trajectories of the stable and unstable manifolds are plotted in the phase space with $u$ on the x-axis and $w$ on the y-axis. The stable manifold is obtained by integrating over a positive time span, while the unstable manifold is obtained by integrating over a negative time span.

**Calculating the Minimum Distance between Stable and Unstable Manifolds**

The minimum distance between the points on the stable manifold and the unstable manifold is determined by calculating the Euclidean distance between each pair of points on the two manifolds. The Euclidean distance between a point $(u_s, w_s)$ on the stable manifold and a point $(u_u, w_u)$ on the unstable manifold is given by:

$$\text{distance} = \sqrt{(u_s - u_u)^2 + (w_s - w_u)^2}. \qquad (16)$$

This computation is performed for each pair of points where $u_s$ and $w_s$ are coordinates of points on the stable manifold and $u_u$ and $w_u$ are coordinates of points on the unstable manifold.

The closest points $(u_{s_{\min}}, w_{s_{\min}})$ and $(u_{u_{\min}}, w_{u_{\min}})$, which correspond to the minimum distance, are found by iterating over all point pairs and keeping track of the smallest distance encountered:

$$\text{min\_distance} = \min_{\forall(u_s, w_s), \forall(u_u, w_u)} \sqrt{(u_s - u_u)^2 + (w_s - w_u)^2}. \qquad (17)$$

Here, min_distance is initialized to infinity, and during the iterations, if a smaller distance is found, min_distance is updated with this new value, and the corresponding points $(u_{s_{\min}}, w_{s_{\min}})$ and $(u_{u_{\min}}, w_{u_{\min}})$ are recorded as the closest points.

### 4.1.2 Discussion of Homo-clinic Orbits

The homoclinic orbit shown in the image is an important concept in the theory of dynamical systems. Homoclinic orbit is a trajectory that intersects an unstable manifold and a stable manifold in phase space, and is known as a phenomenon that shows the complex behavior of nonlinear dynamical systems.

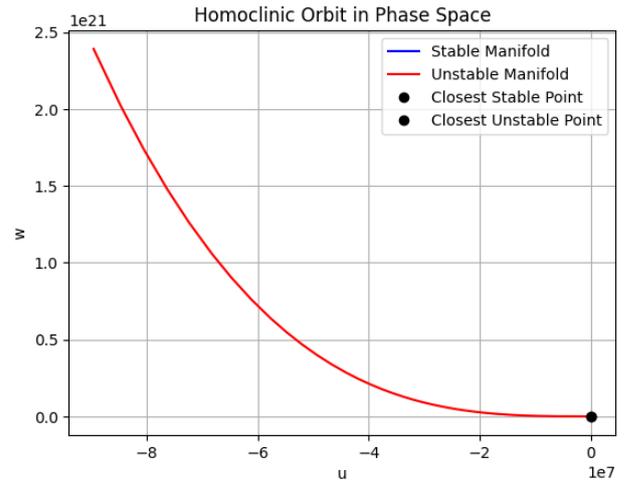

Fig. 19: Homoclinic Orbit in Phase Space

The trajectory appears to depart from a particular point in phase space (perhaps an equilibrium or critical point) and then return to that point again. This is a typical feature of homoclinic orbits, since stable and unstable manifolds intersect at the same point. The fact that the Closest Stable Point and the Closest Unstable Point are very close together on the graph suggests that the system may be approaching chaotic behavior. Such behavior is often analyzed using Melnikov's method, which can show that chaos occurs under certain conditions. When considered in relation to social phenomena, homoclinic orbits suggest that small deviations from a balanced state (equilibrium point) can cause large fluctuations. For example, in an economic system, this is the case when small market fluctuations lead to a major economic crisis. Understanding the dynamics of homoclinic orbits may also apply when social tensions exceed a certain threshold and suddenly riots or revolutions erupt.

Such systems are very sensitive to small external perturbations and changes, which can have unpredictable consequences. Therefore, it is important to carefully manage their dynamics and avoid conditions that lead to extreme behavior in order to maintain system stability.

## Hamiltonian Analysis of a Dynamical System for Homoclinic orbits

We define a Hamiltonian system with parameters $p_1 > 0$ and $s = -1$. The Hamiltonian function $H(u, w)$ for this system is given by the expression:

$$H(u, w) = \frac{w^2}{2} - \frac{p_1 u^2}{2} + \frac{su^4}{4}, \qquad (18)$$

where $u$ and $w$ represent the generalized position and momentum in the phase space, respectively. The parameter $p_1$ is a positive constant that scales the quadratic term of the potential energy, and $s$ is a scaling factor for the quartic term in the potential energy, set to $-1$ in this case.

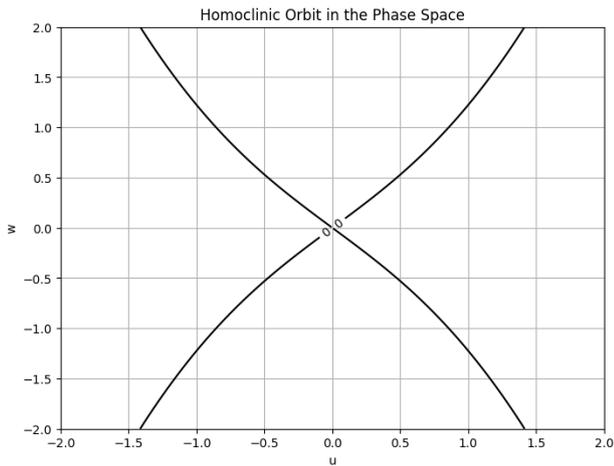

Fig. 20: Homoclinic Orbit in Phase Space

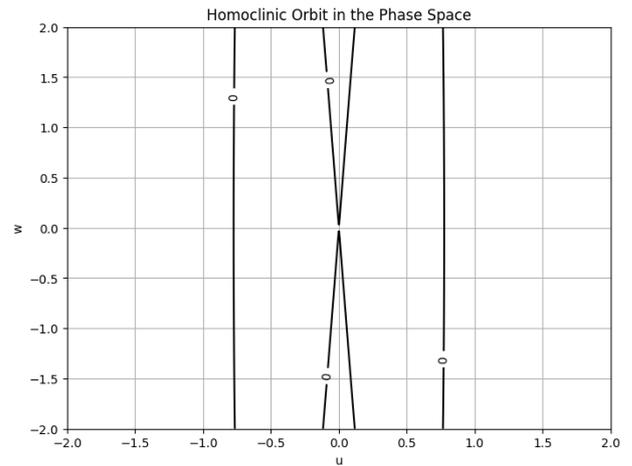

Fig. 21: Homoclinic Orbit in Phase Space

The phase space is visualized by plotting the level curve of the Hamiltonian corresponding to zero energy:

$$H(u, w) = 0. \tag{19}$$

This level curve represents a homoclinic orbit, which is a trajectory in the phase space that joins a saddle point to itself.

The numerical computation involves creating a grid of $u$ and $w$ values and evaluating the Hamiltonian function to obtain the corresponding energy levels. An equipotential contour at zero energy is then plotted, illustrating the homoclinic orbit.

Homoclinic orbit refers to the phenomenon in dynamical systems theory where stable and unstable manifolds of a nonlinear system intersect. This trajectory forms a trajectory in which the system leaves its equilibrium state, called the "saddle point," and returns again. This means that the state of the system is prone to change under certain conditions, and that even small external fluctuations can cause significant changes.

Consideration of homoclinic trajectories in social phenomena shows that even when economic, political, environmental, and other systems appear to be in a long-term equilibrium state, their internal dynamics contain instability, and unexpected events or new information can cause rapid changes. For example, social stress, economic crises, and political change are examples of this type of dynamics.

What can be read from the uploaded graph is that small fluctuations can cause very large changes if the system follows a homoclinic trajectory under certain conditions. This shows how fragile social and economic systems are and how quickly they need to adapt to change.

Applying this to society, we can think, for example, of "revolutionary change," in which a sudden transformation occurs after a long period of political stability, or of "worsening climate change," in which environmental systems change rapidly once a certain threshold is exceeded.

From Figure 21, provided looks like a phase space diagram showing isoclinic orbits. This is a trajectory in phase space starting from and returning to the same saddle point equilibrium. In such a system, the paths represent changes over time in the state of the dynamic system, and the saddle point is an unstable equilibrium.

From a mathematical point of view, the homoclinic trajectory in this figure suggests a nonlinear dynamical system whose motion can be complex and sensitive to initial conditions near the homoclinic loop. When the system is perturbed, it may follow the homoclinic trajectory back to the saddle point or diverge, resulting in chaotic behavior. The presence of multiple homoclinic loops, as suggested by multiple trajectories converging toward the equilibrium point, may indicate a rich and complex dynamic structure that is sensitively dependent on the initial conditions and parameters of the system.

Relating this concept to social phenomena, homoclinic trajectories serve as a metaphor for situations in which a system or society returns to a similar state after experiencing a significant event or disruption. This can be interpreted in the following ways

### 1. Economic cycles

Boom and bust cycles in an economy can resemble the homoclinic behavior of an economy returning to an equilibrium or steady state after a disruption, such as a financial crisis, which can be similar, but not identical, to the pre-crisis state.

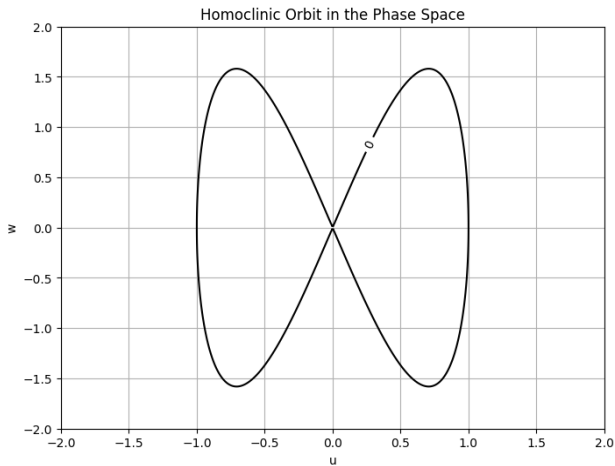

Fig. 22: Homoclinic Orbit in Phase Space

## 2. Social and political change

In a social or political system, a homoclinic trajectory may represent a revolution or reform in which a society experiences major changes but eventually returns to a state that shares characteristics with the initial state. For example, a country may experience political change and then stabilize with a new regime that differs from, but has similarities to, the structure and policies of the previous regime.

## 3. Ecosystem systems

An ecosystem that experiences a disturbance (e.g., forest fire) may eventually grow to a state similar to its original state, showing a homoclinic loop where the final state is similar to the initial state in terms of species composition and ecosystem structure.

## Saddle point or away from the equilibrium state

In this From Figure 23, , the points marked along the isoclinic trajectory may represent specific events or interventions that bring the system back to the saddle point or away from the equilibrium state, depending on the nature and scale of the event. The exact evolution of the system depends on its particular characteristics and the external forces acting on it.

It is important to note that such graphical representations are highly simplified and abstract when applied to real-world social phenomena. Actual social behavior is influenced by a multitude of factors, interactions, and random events that can deviate substantially from the predictable patterns suggested by a two-dimensional state diagram. Therefore, while homoclinic trajectories provide a useful conceptual tool for understanding certain aspects of dynamic behavior, they should be applied with caution rather than interpreted as a literal or comprehensive model of social dynamics.

Homoclinic orbit in a phase space diagram. A homoclinic orbit is a trajectory that starts and ends at the same saddle point in phase space, which is an equilibrium solution to a differential equation.

In terms of social phenomena, homoclinic orbits can serve as a metaphor for situations where events or behaviors depart from a stable state, go through a complex journey, but eventually return to the initial state. This could be likened to economic cycles where economies go through booms and busts but return to some form of equilibrium, or to political systems that experience upheaval before returning to a stable state.

The homoclinic orbit depicted might suggest a system that can absorb disturbances and return to its original state, but the path taken might be complex and unpredictable. This can be seen in many natural and social systems where after some disruption, the system undergoes a series of dynamic changes before settling back into a state of equilibrium.

From the perspective of social dynamics, homoclinic orbits could represent the resilience of social structures or the cyclical nature of certain social behaviors. For example, it might depict the idea that certain societal trends or issues come to the forefront, become intense, and then dissipate, only to potentially re-emerge later. This could apply to social movements, economic policies, cultural trends, and other phenomena that exhibit cyclical patterns.

The complex nature of the trajectory also speaks to the sensitivity of systems to initial conditions. Small changes at the start can have large implications for the trajectory of the system, which can be seen in the context of social systems where initial policy decisions or social actions can lead to very different outcomes.

Please note that while mathematical concepts like homoclinic orbits can provide a useful framework for understanding complex systems, direct analogies to social phenomena should be made with caution.

## Possible transitions between stability and chaos

Homoclinic orbit in a phase space diagram, characterized by trajectories that start and end at the same point, which represents an equilibrium state of the system. In the context of dynamical systems theory, homoclinic orbits indicate the presence of complex dynamics, including possible transitions between stability and chaos. The specific shape and nature of these orbits can provide insights into the resilience of the system and its ability to return to equilibrium after perturbations.

## 1. Economic Systems

Economic cycles can sometimes be represented by homoclinic orbits, where economies experience periods of growth

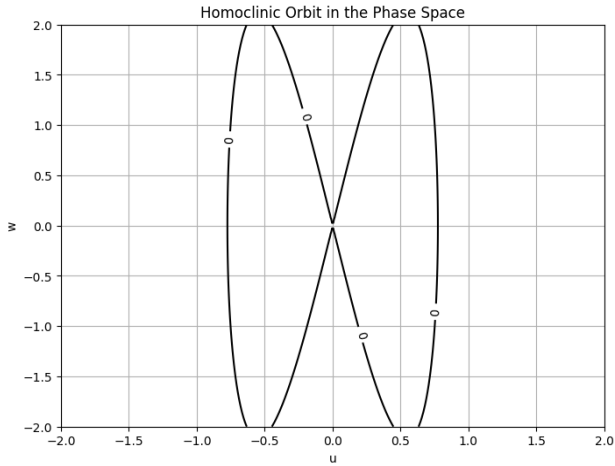

Fig. 23: Homoclinic Orbit in Phase Space

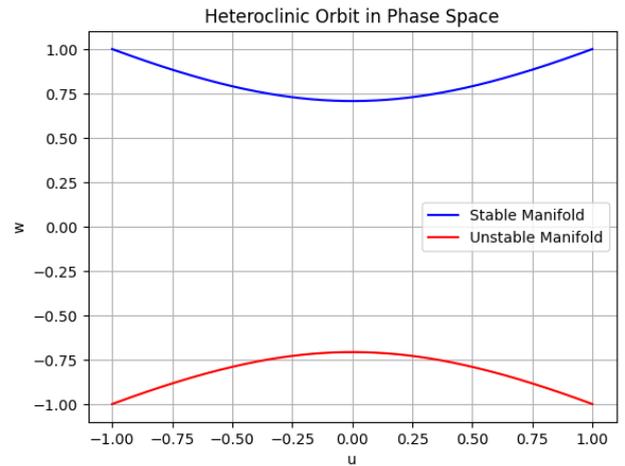

Fig. 24: Heteroclinic Orbits in Phase Space

(boom), followed by a downturn (recession), and eventually a return to some equilibrium state. The specific trajectory can provide insights into how quickly an economy might recover from a shock and whether it may experience lingering effects or return to its original state.

## 2. Political Systems

In politics, a homoclinic orbit could symbolize a political system that goes through a crisis, such as a revolution or upheaval, and then returns to a state of normalcy. The closeness of the trajectories to the equilibrium could represent how radical or conservative changes are over time.

## 3. Cultural Trends

Cultural or societal trends often resurface in new forms after lying dormant. These orbits can represent the resurgence of these trends, evolving each time they reappear but rooted in the same underlying principles or ideas.

In summary, the homoclinic orbits in the provided diagram could be seen as a representation of the cyclical nature of certain social behaviors and the resilience of systems to perturbations. However, these interpretations should be approached with caution, as they are simplifications and can only provide a limited view of the intricate dynamics of real-world systems.

## 4.2 Public opinion and Hetero-clinic Orbits in Dynamical Systems

### 4.2.1 Modeling for Hetero-clinic Orbits

We consider a dynamical system in which the parameters $p_1$ and $\epsilon$ are set such that $p_1 > 0$ and $\epsilon = 0$. Under these conditions, the stable and unstable manifolds can be calculated and visualized in phase space. The manifolds are defined for $u$, which ranges from $-\sqrt{p_1}$ to $\sqrt{p_1}$, as follows:

For the stable manifold:

$$w_{\text{stable}} = \frac{1}{\sqrt{2}}\sqrt{u^2 + p_1}, \tag{20}$$

and for the unstable manifold:

$$w_{\text{unstable}} = -\frac{1}{\sqrt{2}}\sqrt{u^2 + p_1}. \tag{21}$$

These equations describe the shape of the manifolds in the $(u, w)$ phase space for the given dynamical system. The stable manifold is associated with the positive square root and represents trajectories that converge to a fixed point, while the unstable manifold is associated with the negative square root and represents trajectories that diverge from a fixed point.

These manifolds provides insight into the structure of the phase space and the behavior of the system near these manifolds.

### 4.2.2 Discussion of Hetero-clinic Orbits

The heteroclinic orbits in the image show the path of the system's solution between different equilibrium points. Stable manifolds (blue lines) represent the tendency of the system to gravitate toward some equilibrium state, while unstable manifolds (red lines) represent the tendency of the system to leave some equilibrium state. Heteroclinic orbits suggest where these manifolds intersect, i.e., transitions from one equilibrium state to another are possible.

Such dynamic behavior is of great interest when modeling sudden changes or abrupt transitions from one state to another in a social or economic context. For example, the transition from political stability to instability or from rapid market growth to contraction can be addressed.

In social phenomena, heteroclinic orbits often show a tendency toward large fluctuations and u nexpected events. For example, a quiet, stable society is suddenly hit by political upheaval, or an economy rapidly collapses. These changes are often difficult to predict, and small fluctuations can trigger major system-wide changes.

Mathematical models play an important role in explaining social phenomena, helping us understand their predictability and difficulty of co ntrol. Concepts such as heteroclinic trajectories may be useful to better understand how systems move from one state to another.

## Aknowlegement

The author is grateful for discussion with Prof. Serge Galam and Prof.Akira Ishii. This research is supported by Grant-in-Aid for Scientific Research Project FY 2019-2021, Research Project/Area No. 19K04881, "Construction of a new theory of opinion dynamics that can describe the real picture of society by introducing trust and distrust".

## References

[Strogatz(1994)] Strogatz, S. H. "Nonlinear Dynamics and Chaos: With Applications to Physics, Biology, Chemistry, and Engineering." Westview Press, **1994**.

[Guckenheimer & Holmes(1983)] Guckenheimer, J., & Holmes, P. "Nonlinear Oscillations, Dynamical Systems, and Bifurcations of Vector Fields." Springer-Verlag, **1983**.

[Kuznetsov(2004)] Kuznetsov, Y. A. "Elements of Applied Bifurcation Theory." Springer, **2004**.

[Wiggins(2003)] Wiggins, S. "Introduction to Applied Nonlinear Dynamical Systems and Chaos." Springer, **2003**.

[Marsden & McCracken(1976)] Marsden, J. E., & McCracken, M. "The Hopf Bifurcation and Its Applications." Springer-Verlag, **1976**.

[Seydel(2010)] Seydel, R. "Practical Bifurcation and Stability Analysis." Springer, **2010**.

[Scheffer et al.(2009)] Scheffer, M., et al. "Early-warning signals for critical transitions." *Nature*, vol. 461, pp. 53-59, **2009**.

[Gladwell(2000)] Gladwell, M. "The Tipping Point: How Little Things Can Make a Big Difference." Little, Brown and Company, **2000**.

[Epstein(2006)] Epstein, J. M. "Generative Social Science: Studies in Agent-Based Computational Modeling." Princeton University Press, **2006**.

[Castellano et al.(2009)] Castellano, C., Fortunato, S., & Loreto, V. "Statistical physics of social dynamics." *Reviews of Modern Physics*, vol. 81, no. 2, pp. 591, **2009**.

[Granovetter(1978)] Granovetter, M. "Threshold Models of Collective Behavior." *American Journal of Sociology*, vol. 83, no. 6, pp. 1420-1443, **1978**.

[Lenton et al.(2008)] Lenton, T. M., et al. "Tipping elements in the Earth's climate system." *Proceedings of the National Academy of Sciences*, vol. 105, no. 6, pp. 1786-1793, **2008**.

[Tainter(1988)] Tainter, J. A. "The Collapse of Complex Societies." Cambridge University Press, **1988**.

[Kuran(1991)] Kuran, T. "Now out of never: The element of surprise in the East European revolution of 1989." *World Politics*, vol. 44, no. 1, pp. 7-48, **1991**.

[Granovetter(1978)] Granovetter, M. "Threshold models of collective behavior." *American Journal of Sociology*, vol. 83, no. 6, pp. 1420-1443, **1978**.

[Epstein(2002)] Epstein, J. M. "Modeling Civil Violence: An Agent-Based Computational Approach." *Proceedings of the National Academy of Sciences*, vol. 99, Suppl 3, pp. 7243-7250, **2002**.

[Centola & Macy(2007)] Centola, D., & Macy, M. "Complex Contagions and the Weakness of Long Ties." *American Journal of Sociology*, vol. 113, no. 3, pp. 702-734, **2007**.

[Melnikov(1963)] Melnikov, V. K. On the stability of the center for time periodic perturbations. *Trans. Moscow Math. Soc.* **1963**, *12*, 1–57.

[Guckenheimer & Holmes(1983)] Guckenheimer, J.; Holmes, P. *Nonlinear Oscillations, Dynamical Systems, and Bifurcations of Vector Fields*. Springer: **1983**.

[Wiggins(1990)] Wiggins, S. *Introduction to Applied Nonlinear Dynamical Systems and Chaos*. Springer: **1990**.

[Kuznetsov(1995)] Kuznetsov, Y. A. *Elements of Applied Bifurcation Theory*. Springer: **1995**.

[Brock & Hommes(1997)] Brock, W. A.; Hommes, C. H. A rational route to randomness. *Econometrica* **1997**, *65*(5), 1059–1095.

[Farmer(1999)] Doyne Farmer, J. Physicists attempt to scale the ivory towers of finance. *Computing in Science & Engineering* **1999**, *1*(6), 26–39.

[Hastings et al.(1993)] Hastings, A.; Hom, C. L.; Ellner, S.; Turchin, P.; Godfray, H. C. J. Chaos in Ecology: Is Mother Nature a Strange Attractor? *Annual Review of Ecology and Systematics* **1993**, *24*, 1–33.

[Moon(1987)] Moon, F. C. *Chaotic Vibrations: An Introduction for Applied Scientists and Engineers*. Wiley: 1987.

[Carr(1981)] Carr, J.. Applications of Center Manifold Theory. Springer, **1981**.

[Guckenheimer(1983)] Guckenheimer, J. and Holmes, P. Nonlinear Oscillations, Dynamical Systems, and Bifurcations of Vector Fields. Springer, **1983**.

[Wiggins(2003)] Wiggins, S.. Introduction to Applied Nonlinear Dynamical Systems and Chaos. Springer, 2nd ed., **2003**.

[Chen2002] Chen, L. Q. (2002). Phase-field models for microstructure evolution. *Annual Review of Materials Research*, *32*(1), 113-140, **2002**.

[Boettinger2002] Boettinger, W. J., Warren, J. A., Beckermann, C., & Karma, A. (2002). Phase-field simulation of solidification. *Annual Review of Materials Research*, *32*(1), 163-194, **2002**.

[Hegselmann & Krause(2002)] Hegselmann, R., & Krause, U. Opinion Dynamics and Bounded Confidence Models, Analysis, and Simulation. *Journal of Artificial Society and Social Simulation* **2002**, *5*, 1-33.

[Ishii & Kawahata(2018)] Ishii A. & Kawahata, Y. Opinion Dynamics Theory for Analysis of Consensus Formation and Division of Opinion on the Internet. In: Proceedings of The 22nd Asia Pacific Symposium on Intelligent and

Evolutionary Systems, 71-76, **2018**. arXiv:1812.11845 [physics.soc-ph]

[Ishii(2019)] Ishii A. Opinion Dynamics Theory Considering Trust and Suspicion in Human Relations. In: Morais D., Carreras A., de Almeida A., Vetschera R. (eds) Group Decision and Negotiation: Behavior, Models, and Support. GDN 2019. *Lecture Notes in Business Information Processing* 351, Springer, Cham 193-204, **2019**.

[Ishii & Kawahata(2019)] Ishii A. & Kawahata, Y. Opinion dynamics theory considering interpersonal relationship of trust and distrust and media effects. In: The 33rd Annual Conference of the Japanese Society for Artificial Intelligence 33. JSAI2019 2F3-OS-5a-05, **2019**.

[Agarwal et al.(2011)] Agarwal, A, Xie, B., Vovsha, I., Rambow, O. & Passonneau, R. Sentiment analysis of twitter data. In: Proceedings of the workshop on languages in social media. Association for Computational Linguistics 30-38, **2011**.

[Siersdorfer et al.(2010)] Siersdorfer, S., Chelaru, S. & Nejdl, W. How useful are your comments?: analyzing and predicting youtube comments and comment ratings. In: Proceedings of the 19th international conference on World wide web. 891-900, **2010**.

[Wilson et al.(2005)] Wilson, T., Wiebe, J., & Hoffmann, P. Recognizing contextual polarity in phrase-level sentiment analysis. In: Proceedings of the conference on human language technology and empirical methods in natural language processing 347-354, **2005**.

[Sasahara et al.(2020)] Sasahara, H., Chen, W., Peng, H., Ciampaglia, G. L., Flammini, A. & Menczer, F. On the Inevitability of Online Echo Chambers. arXiv: 1905.03919v2, **2020**.

[Ishii and Kawahata(2018)] Ishii, A.; Kawahata, Y. Opinion Dynamics Theory for Analysis of Consensus Formation and Division of Opinion on the Internet. In *Proceedings of The 22nd Asia Pacific Symposium on Intelligent and Evolutionary Systems (IES2018)*, 71-76; arXiv:1812.11845 [physics.soc-ph], **2018**.

[Ishii(2019)] Ishii, A. Opinion Dynamics Theory Considering Trust and Suspicion in Human Relations. In *Group Decision and Negotiation: Behavior, Models, and Support. GDN 2019. Lecture Notes in Business Information Processing*, Morais, D.; Carreras, A.; de Almeida, A.; Vetschera, R. (eds), **2019**, *351*, 193-204.

[Ishii and Kawahata(2019)] Ishii, A.; Kawahata, Y. Opinion dynamics theory considering interpersonal relationship of trust and distrust and media effects. In *The 33rd Annual Conference of the Japanese Society for Artificial Intelligence*, JSAI2019 2F3-OS-5a-05, **2019**.

[Okano and Ishii(2019)] Okano, N.; Ishii, A. Isolated, untrusted people in society and charismatic person using opinion dynamics. In *Proceedings of ABCSS2019 in Web Intelligence 2019*, 1-6, **2019**.

[Ishii and Kawahata(2019)] Ishii, A.; Kawahata, Y. New Opinion dynamics theory considering interpersonal relationship of both trust and distrust. In *Proceedings of ABCSS2019 in Web Intelligence 2019*, 43-50, **2019**.

[Okano and Ishii(2019)] Okano, N.; Ishii, A. Sociophysics approach of simulation of charismatic person and distrusted people in society using opinion dynamics. In *Proceedings of the 23rd Asia-Pacific Symposium on Intelligent and Evolutionary Systems*, 238-252, **2019**.

[Ishii and Okano(2021)] Ishii, A, and Nozomi,O. Sociophysics approach of simulation of mass media effects in society using new opinion dynamics. In *Intelligent Systems and Applications: Proceedings of the 2020 Intelligent Systems Conference (IntelliSys) Volume 3*. Springer International Publishing, **2021**.

[Ishii and Kawahata(2020)] Ishii, A.; Kawahata, Y. Theory of opinion distribution in human relations where trust and distrust mixed. In Czarnowski, I., et al. (eds.), *Intelligent Decision Technologies*, Smart Innovation, Systems and Technologies 193, **2020**; pp. 471-478.

[Ishii et al.(2021)] Ishii, A.; Okano, N.; Nishikawa, M. Social Simulation of Intergroup Conflicts Using a New Model of Opinion Dynamics. *Front. Phys.* **2021**, 9:640925. doi: 10.3389/fphy.2021.640925.

[Ishii et al.(2020)] Ishii, A.; Yomura, I.; Okano, N. Opinion Dynamics Including both Trust and Distrust in Human Relation for Various Network Structure. In *The Proceeding of TAAI 2020*, in press, **2020**.

[Fujii and Ishii(2020)] Fujii, M.; Ishii, A. The simulation of diffusion of innovations using new opinion dynamics. In *The 2020 IEEE/WIC/ACM International Joint Conference on Web Intelligence and Intelligent Agent Technology*, in press, **2020**.

[Ishii & Okano(2021)] Ishii, A, Okano, N. Social Simulation of a Divided Society Using Opinion Dynamics. *Proceedings of the 2020 IEEE/WIC/ACM International Joint Conference on Web Intelligence and Intelligent Agent Technology* is press, **2021**.

[Ishii & Okano(2021)] Ishii, A., & Okano, N. Sociophysics Approach of Simulation of Mass Media Effects in Society Using New Opinion Dynamics. In *Intelligent Systems and Applications (Proceedings of the 2020 Intelligent Systems Conference (IntelliSys) Volume 3)*, pp. 13-28. Springer, **2021**.

[Okano & Ishii(2021)] Okano, N. & Ishii, A. Opinion dynamics on a dual network of neighbor relations and society as a whole using the Trust-Distrust model. In *Springer Nature - Book Series: Transactions on Computational Science & Computational Intelligence (The 23rd International Conference on Artificial Intelligence (ICAI'21))*, **2021**.